\documentclass[epjST]{svjour}
\usepackage{doi}
\usepackage{hyperref}
\hypersetup{
 colorlinks=true,
 linkcolor=blue,
 filecolor=magenta, 
 urlcolor=cyan,
 pdftitle={Catalysis of Fusion by Heavy Nuclei},
 }
\usepackage{amsmath}
\usepackage{amssymb}
\usepackage{graphicx}
\usepackage{float}

\usepackage{mathtools} 
\usepackage{xfrac} 
\usepackage{ulem}

\newcommand{\req}[1]{Eq.~({\ref{#1}})}

\newcommand{\reff}[1]{Fig.~{\ref{#1}}}

\newcommand{\orcJR}{0000-0001-8217-1484}
\newcommand{\orcCMG}{0000-0001-9985-1822}
\newcommand{\orcidicon}{\includegraphics[width=0.32cm]{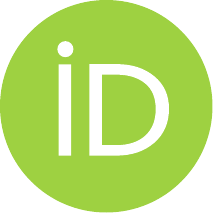}}
\newcommand{\orc}[1]{\href{https://orcid.org/#1}{\orcidicon}}

\begin{document} 

%\title{Heavy nuclear catalysts for screening enhancement of fusion reactions}
\title{Nuclear Fusion Enhancement by Heavy Nuclear Catalysts}
%\shortauthors{Grayson et al.}

\author{Christopher Grayson${}^{1}$\orc{\orcCMG}%
%\thanks{Corresponding author \email{chrisgray1044@gmail.com}}%
, Johann Rafelski${}^{2}$\orc{\orcJR}%
%\thanks{Corresponding author \email{johannr@arizona.edu}}%
\institute{${}^1$Wigner Research Centre for Physics, 1121 Budapest, Hungary\\
${}^2$Department of Physics, The University of Arizona, Tucson, Arizona 85721, USA }
}

\abstract{%
We seek to understand the effect of high electron density in the proximity of a heavy nucleus on the fusion reaction rates in a hot plasma phase. We investigate quantitatively the catalytic effect of gold ($Z=79$) ions embedded in an electron plasma created due to plasmonic focusing of high-intensity short laser pulses. Using self-consistent strong plasma screening, we find highly significant changes in the internuclear potential of light elements present nearby. For gold, we see a $14\,$keV change in the internuclear potential near the nuclear surface, independent of the long-distance thermal Debye-H\"uckel screening. The dense polarization cloud of electrons around the gold catalyst leads to a $\sim 1.5$ enhancement of proton-boron ($^{11}$B) fusion above $T=100$\,keV.
}

\maketitle

\section{Introduction}\label{sec:intro}

This article expands on the work in \cite{Grayson:2024uwg} relating it to fusion enhancement in gold nanoparticle-doped fusion targets. In the earlier work~\cite{Grayson:2024uwg}, the effect of self-consistent screening of light nuclei during Big Bang nucleosynthesis (BBN) was found to lead to a small correction for thermonuclear reactions but to have a very strong dependence on nuclear charge $Ze$. This article investigates self-consistent strong screening of ``heavy nuclear catalyst'' or high $Z$-ionic impurities in a plasma. 

Heavy nuclear catalysts, such as gold ($Z=79$), generate a dense local screening cloud of electrons within a hot plasma. The Coulomb barrier is significantly reduced within the electron cloud, and lighter nuclei can more easily participate in thermonuclear reactions. Dispersed gold ions are often present in plasma generated in laser-fusion experiments using gold film \cite{PhysRevE.71.036412,PhysRevLett.114.124801}, in the hohlraum used in inertial confinement fusion experiments \cite{hurricane2014fuel,casey2023towards}, and introduced by plasmoinc gold nanoparticles \cite{vallieres2021enhanced,PRXEnergy.1.023001,universe9050233,csernai2023crater,csernai2024femtoscopy,szokol2024pulsedlaserintensitydependence,papp2024pics,kroo2024indication,Csernai:2023vzf}.
In these plasmonic nanoparticle experiments, a huge number of conducting electrons oscillate with the electromagnetic field of an ultrafast pulse laser in a localized surface plasmonic resonance (LSPR). The screening enhancement of fusion reactions could occur locally within the cloud of conducting electrons around these nanorods. The electron cloud oscillation in an LSPR is a coherent quantum process. We explore this concept semi-classically by analogy using a high-temperature electron-ion plasma around an ionized gold nucleus.
\begin{figure}%%[h!]
 \centering
 \includegraphics[width=0.3\columnwidth]{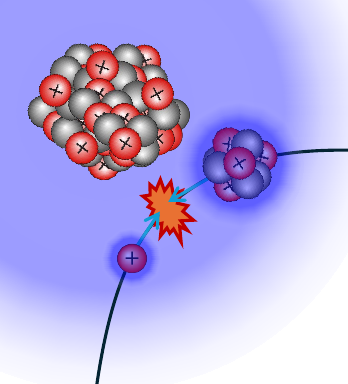}
 \caption{Depiction of heavy nuclei catalyzing fusion reactions. Two particles (proton and boron) collide in the presence of a large screening cloud of electrons generated by a stationary heavy nucleus such as Au ($Z=79$).}
 \label{fig:cartoon}
\end{figure}

Self-consistent strong screening refers to solving the Poison equation, including the induced charge density, without any approximation in its dependence on the electric potential $\phi$. In this case, the charge density is a nonlinear function of the solution potential $\phi$. Electrons surround an ion's charge $Ze$ (elementary charge $e>0$), which reduces the electromagnetic influence of other nuclear charges at large distances. In the context of nuclear reactions, this reduction in electrostatic repulsion reduces the Coulomb barrier and facilitates an increased penetration probability. This implies that electromagnetic screening increases thermonuclear reaction rates \cite{Salpeter:1954nc}. Heavy nuclear catalysts are high Z nuclei, which provide a large screening cloud with which lighter nuclei can enter and more easily undergo thermonuclear reactions. This process is depicted in \reff{fig:cartoon} where two light nuclei collide in the screening cloud of a third stationary heavy nucleus ($m_\text{Au} \gg T$), which acts as a catalyst for the reaction.

Often plasma studies assume the ``weak-field'' limit where the electromagnetic potential energy $q\phi(r)$ is small compared to the thermal energy $T$
\begin{equation}\label{eq:weakcond}
 \frac{q \phi(r)}{T} \ll 1\,,
\end{equation}
where we set Boltzmann's constant $k_B = 1$. Plasmas satisfying this condition are weakly coupled, and plasma effects lead to linear corrections to the potential, as in Debye-H\"{u}ckel theory~\cite{Debye:1923}. 

Even if a plasma is weakly coupled, nonlinear corrections to the potential may be relevant in thermonuclear reactions. This is because the Gamow energy $E_G$, the energy at most reactions occur, is higher than thermal energy~\cite{Shaviv:1996} and probes the short-distance electromagnetic potential. The inter-nuclear distance corresponding to Gamow energy is typically on the order of femtometers, such that the weak screening condition is not satisfied for the short-range potential relevant to the Coulomb barrier
\begin{figure}
 \centering
 \includegraphics[width=0.7\columnwidth]{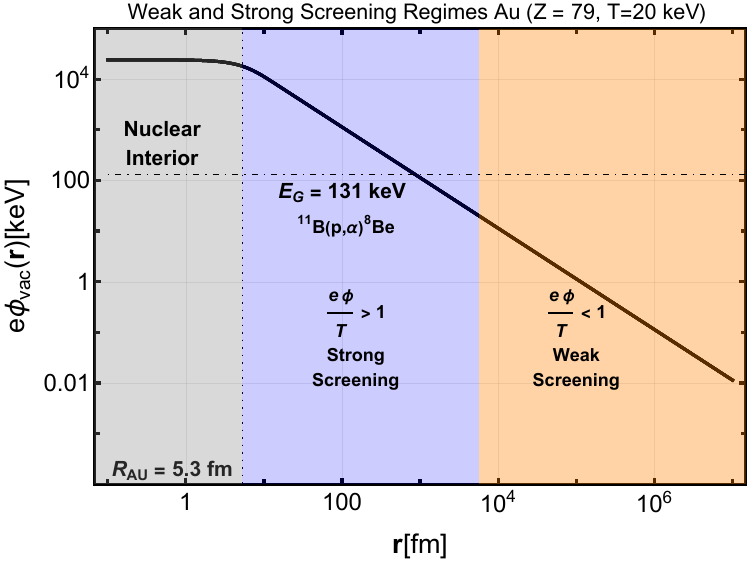}
 \caption{We show the ratio of the electromagnetic potential energy to the thermal energy for the vacuum potential of gold nuclei. The portion of the potential with strong screening corrections is shown in purple, and the weak screening region is shown in orange. The Gamow energy \req{eq:gamow} for a proton boron reaction probes the strong screening regime}
 \label{fig:diagram}
\end{figure}
\begin{equation}\label{eq:strongcond}
 \frac{Ze \phi(r_{E_G})}{T} > 1\,,
\end{equation}
where $r_{E_G}$ is the classical turning point at the Gamow energy. 
\begin{equation}\label{eq:gamow}
 E_G = \left(\frac{(\pi T Z_1 Z_2 \alpha)^2 \mu_r }{2}\right)^{1/3}\,.
\end{equation}
The reduced mass of the colliding light nuclei is $\mu_r$, and $Z_1$ and $Z_2$ are the respective charges of the nuclei. As depicted in \reff{fig:diagram}, the condition \req{eq:strongcond} implies that although a plasma may be treated as weakly coupled at large distances, locally, the short-range potential can have nonlinear corrections simply because the electrostatic energy close to a nucleus can be much higher than the thermal energy. 

The self-consistent strong polarization potential is described by the Poisson-Boltzmann equation~\cite{Dzitko:1995xyz,Gruzinov:1998,Bruggen:1997,Bruggen:2000,Bi:2000,Liolios:2004,Luo:2020,Grayson:2024uwg}. We solve the Poisson-Boltzmann equation directly by implementing a finite-sized source charge density and using Fermi-Dirac statistics. This description of screening reproduces the Thomas-Fermi model in the zero temperature limit and Debye-H\"uckel screening in the high temperature $\phi/T\ll 1$ limit. This allows us to simultaneously describe bound electrons and outer electrons, which join the thermal screening cloud in the plasma at various temperatures. The Poisson Boltzmann equation is also relevant to describe the sheath potential in Target Normal Sheath Acceleration (TNSA)~\cite{PhysRevLett.90.185002,bochkarev2006investigation}.

Self-consistent strong screening differs from other `strong screening' models ~\cite{Salpeter:1954nc,Dewitt1973,Itoh1977,Itoh1979,Ichimaru:1982,Chugunov:2007ae,Chugunov:2009ic,Kravchuk:2014xsa} developed for white dwarfs which are applicable for plasmas where the Coulomb energy is much larger than the thermal energy. Models studying strong screening are typically interested in plasmas with a Coulomb coupling parameter
\begin{equation}
 \Gamma_e = \frac{Z_1 Z_2 \alpha \hbar c}{T a_e}, \quad \text{with} \quad a_e = \left(\frac{3}{4 \pi n_e}\right)^{1/3}\,,
\end{equation}
which is $1 \leq\Gamma_e \leq 200$. In this work, we study `intermediate screening,' which we call self-consistent strong screening since it captures both the strong and weak field regimes.

Section~\ref{NSLN} reviews the theoretical background of self-consistent strong screening. Section~\ref{sec:Solve} presents numerical solutions to the Poisson-Boltzmann equation for heavy Au particles in a laser plasma. Section~\ref{sec:Enhancement} estimated the reaction rate enhancement for proton boron fusion. Section~\ref{sec:Conc_Disc} reviews our results and discusses their implications for fusion.

\section{Self-consistent strong screening of heavy nuclear catalysts}\label{NSLN}

We find the total potential $\phi(r) = \phi_\text{ext}(r)+\phi_\text{ind}(r)$ in plasma by solving the Poisson equation with the induced polarization charge density $\rho_\mathrm{ind}(r)$ and the external charge density of the light nuclei $\rho_\mathrm{ext}(r)$
\begin{equation}\label{eq:Poss}
 -\nabla^2\phi(r) =\rho_\mathrm{tot}(r)/\varepsilon_0= [\rho_{\mathrm{ext}}(r) + \rho_{\mathrm{ind}}(r)]/\varepsilon_0\,,
\end{equation}
where $\varepsilon_0$ is the vacuum permittivity. The equilibrium-induced charge density is the difference between the charge density of electrons and protons
\begin{equation}\label{eq:chargesum}
 \rho_{\mathrm{ind}}(r) = e n_\text{p}(r) - e n_-(r) \approx - e n_-(r) \,,
\end{equation}
where $n_-(r)$ represents the number density of electrons and $n_\text{p}(r)$ of protons. We study the electron plasma at keV energies, which forms from the laser interaction with matter for intensities of $10^{17}-10^{20}$ W/cm$^2$ on timescales of 20 fs. On this time scale, we assume electrons in the plasma are approximately at thermal equilibrium and ions have not yet reacted since they are not mobile enough to participate in screening due to their much larger mass. Including plasma dynamics in the self-consistent strong screening is the subject of future work.

We introduce the unit-less potential
\begin{equation}\label{eq:indchgB}
 \Phi(r) \equiv \frac{e\phi(r)}{T}\,,
\end{equation}
which compares the Coulomb energy $e\phi$ to the plasma temperature $T$. 
Similarly we introduce the re-scaled external charge distribution $P_{\mathrm{ext}}$
\begin{equation}\label{eq:Pext}
 P_{\mathrm{ext}}(r) \equiv e \frac{\rho_{\mathrm{ext}}(r)}{\varepsilon_0 T} = \frac{4 \pi Z \alpha \hbar c}{\pi^{3/2}R^3T}e^{-\frac{r^2}{R^2}}\,,
\end{equation}
we chose to model the charge distribution of a nucleus as a Gaussian to make self-consistent screening soluble. Here $R$ is the root-mean-squared charge radius
\begin{equation}
 R = \sqrt{\frac{2}{3}} R_\text{Au}\,.
\end{equation}
For the radius of an Au (gold) atom, we use the charge radius $R_\text{Au} = 5.3\,$fm~\cite{DeVries:1987atn}. 
We rewrite the Poisson equation \req{eq:Poss} using the effective screening mass defined in \cite{Grayson:2024uwg}
\begin{equation}\label{eq:mscreen}
\frac{m_s^2(\Phi)}{(\hbar c)^2} \equiv -\frac{ e\rho_{\mathrm{ind}}(r)/(\varepsilon_0 T)}{ \Phi(r)}\,. 
\end{equation}
The screening mass $m_s^2$ is related to the usual Deybe screening mass in the weak screening limit
 \begin{equation}\label{eq:debye}
 \frac{m_s^2(\Phi \ll 1)}{(\hbar c)^2} \approx \frac{m_D^2}{(\hbar c)^2} =\frac{1}{\lambda_D^2} = \frac{4 \pi \alpha \hbar c}{T} \, n_{\mathrm{eq}}\,.
\end{equation}
but exhibits additional nonlinear behavior due to its dependence on the potential.
Using this screening mass \req{eq:Poss} takes a form which is analogous to Debye-H\"uckel theory
\begin{equation}\label{eq:PossBoltz}
 -\nabla^2\Phi(r) + \frac{m_s^2(\Phi)}{(\hbar c)^2}\Phi(r) = P_{\mathrm{ext}}(r)\,,
\end{equation}
Which is then solved numerically for $\Phi(r)$. We describe the screening as using the relativistic Fermi-Dirac distribution \cite{Grayson:2024uwg}
\begin{equation}\label{eq:Fermi}
 f_F^\pm(x,p) = \frac{1}{\exp\left[u_{\mu}(p^{\mu}+q A^{\mu}(x))\right]/T+1}\,.
\end{equation}
where $q = \pm e$ is the charge of the fermions, $u_{\mu} = (1,0,0,0)$ is the global velocity of the plasma, and $A^{\mu}$ is the 4-potential in the plasma. The screening mass for the Fermi-Dirac distribution for electrons is
\begin{equation}\label{eq:indchgF}
 m^2_{s,\mathrm{FD}}(\Phi) = \frac{4 \pi \alpha }{ T \Phi (r)} \int \frac{d^3\boldsymbol{p}}{(2 \pi)^3 } \frac{1}{\exp\left(p_0/T-\Phi(r) \right)+1}\,,
\end{equation}
where $p_0=\sqrt{|\mathbf{p}|+m^2}$. In the limit, $T\xrightarrow{}0$ \req{eq:indchgF} returns the usual relativistic Thomas-Fermi model. For convenience, we use the electron-positron plasma Debye mass from \cite{Grayson:2024uwg} removing the factor of 2 accounting for the two species, which matches \req{eq:indchgF} for $\phi/T>1$. 
\begin{equation}\label{eq:indchgF2}
 m^2_{s,\mathrm{FD}}(\Phi) \approx \frac{4 \pi \alpha }{ T \Phi (r)} \int \frac{d^3\boldsymbol{p}}{(2 \pi)^3 } \frac{\sinh\left[\Phi(r) \right]}{\cosh\left(p_0/T\right)+\cosh\left[\Phi(r) \right]}\,,
\end{equation}
this ensures the plasma is neutral at large distances, which is not true in \req{eq:indchgF} where only electrons are considered. In a more exact model, we would use a chemical potential to specify the distribution of electrons and protons such that the plasma is neutral. However, since the Debye mass is calculated in the static limit, it is not sensitive to the mass difference of electrons and ions, only the density of charges, which would, in principle, be the same. It would be more important to fix the density to those found in electron plasmas formed during laser interactions with matter, but this is the subject of future work.

\section{Solving the strong field Poisson equation self consistently}\label{sec:Solve}

The numerical solution method to \req{eq:PossBoltz} is described in \cite{Grayson:2024uwg}. To summarize, we start the integration of \req{eq:PossBoltz} at large values of $r$ where the weak field condition \req{eq:weakcond} applies since there once can find an analytic Debye type solution, up to an overall normalization due to the additional screening that will occur at short ranges.
One can integrate inwards with any standard solver to small values of $r$ varying a shooting parameter $Q_{\mathrm{eff}}$ until the solution converges sufficiently close to $r=0$.

\begin{figure}%[h]
 \centering
\includegraphics[width=0.9\columnwidth]{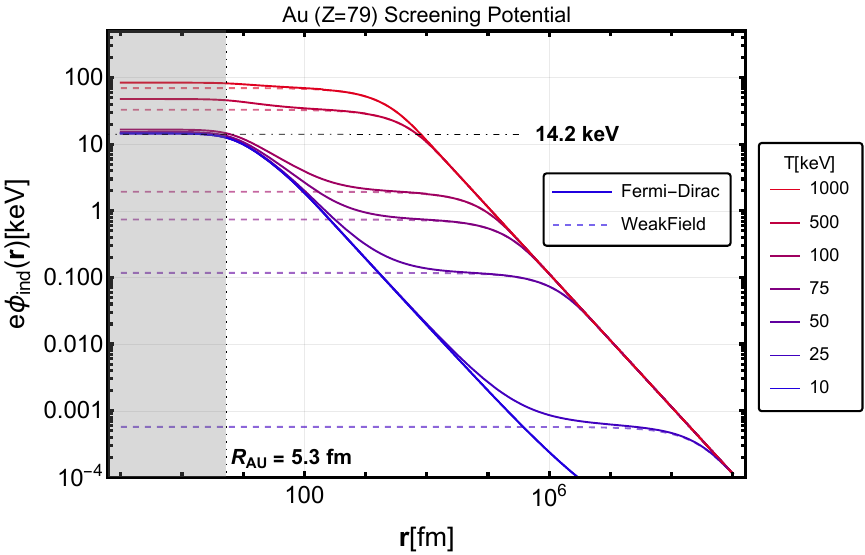}
 \caption{The potential $e\phi_{\text{ind}}$ due to the induced screening charge density for Fermi-Dirac self-consistent strong screening \req{eq:Fermi} at various temperatures is shown as solid lines ranging from blue to red. The weak polarization potential is shown as dashed lines ranging from blue to red. Overall screening decreases with temperature $T$, but the difference between weak and strong becomes larger for small $T$. The gray area shows the nuclear interior at a radius of $R_\text{Au} = 5.3\,$fm, where the nuclear potential would take over.}
 \label{fig:pot_Comp}
\end{figure}

We are interested in the potential due to screening charges $\phi_\text{ind}$ where
\begin{equation}
\phi_\text{ind}(r)\equiv \phi(r)-\phi_{\text{vac}}(r)\,,
\end{equation} 
with $\phi$ being the total potential in the plasma and $\phi_{\text{vac}}$ the potential in vacuum. The induced polarization potential $\phi_\text{ind}(0)$ is the potential due to the screening electron cloud and is responsible for the reduction pf electrostatic potential energy between charges. The potential difference $\phi_\text{ind}(0)$ is proportional to the change in the internuclear potential at the origin, often denoted as $H(0)$, which is related to screening enhancement of reaction rates $\mathcal{F}_{\text{sc}}$ \cite{Salpeter:1954nc} 
\begin{equation}\label{eq:weakenhance}
 \mathcal{F}_{\text{sc}} 
 \sim \exp{\left( \frac{H(0)}
 {T}\right)} \sim \exp{\left(\frac{ e\phi_\text{ind}(0)}{T}\right)} \, .
\end{equation}
The quantity $\phi_\text{ind}(r)$ is plotted for a gold ion in \reff{fig:pot_Comp}. Self-consistent strong screening predicts a larger induced polarization potential than weak screening due to its nonlinear dependence on the potential, which is exponential in the Boltzmann limit~\cite{Grayson:2024uwg}. This effect is more pronounced near the origin where $e\phi/T$ is large. The dashed lines show the weak polarization potential, which is nearly constant up to distances $\lambda_D$ and equal to its value at the origin
\begin{equation}\label{eq:weakorg}
 e\phi_\text{ind}(0) = Z \alpha m_D\,.
\end{equation}
In \reff{fig:pot_Comp}, the overlapping inner blue curves show the potential due to electrons bound to gold, while the outer red curve shows the potential due to the thermal screening cloud. This is because \req{eq:PossBoltz} reduces to Thomas-Fermi Theory in the zero temperature limit and returns Debye-H\"ukel screening in the large distance limit $e\phi(\infty)/T \ll 1$. We can interpret the transition from the inner set of blue curves to the parallel larger distance red curves as outer electrons joining the Debye-H\"uckel type thermal screening cloud at increasing temperatures.

In \reff{fig:pot_Comp}, we see that the self-consistent strong screening contribution to the potential near the origin is additive to the weak screening contribution \req{eq:weakorg}. By subtracting the weak screening contributions, shown as dashed lines, from the full solution at each temperature, we find the strong screening contribution for gold (Z=79) independent of temperature around $14.2$\,keV. This should be close to the electric potential shift due to the electron cloud around a gold atom at zero temperature. Using the numerical fit for the shift in the potential from Hartree-Fock wavefunctions \cite{Garrett1967PotentialES}, and estimating the Thomas-Fermi constant $a_{\text{TF}}$ from the tabulated values in \cite{Garrett1967PotentialES} for $Z=79$, $a\approx 1.598$, we find
\begin{equation}
 e\phi_\text{HF}(0) = \alpha^2 Z^{4/3}a_{\text{TF}}m_e= 14.7\,\text{keV}\,.
\end{equation}
This is very close to our result on the keV scale. When including electron exchange corrections, we find a closer value to this estimate, $14.5 keV$, but the hundreds of eV difference would be negligible for tunneling in a thermonuclear reaction.

Ionized light nuclei with $Z=1$ and plasma thermal energy $\sim \frac{3}{2} T$ would penetrate the screening cloud of gold up to $R_\text{Au}$ above $10\,$keV. Light nuclei would react in a screened environment at this distance with a considerably reduced internuclear potential $\sim 14$\,keV, enhancing fusion reaction rates as illustrated in \reff{fig:diagram}. 

In principle, one might expect that due to the nonlinear form of \req{eq:PossBoltz}, the value of the strong and weak polarization potential would not be additive. However, as discussed in \cite{Grayson:2024uwg} the self-consistent polarization potential at the origin is related to the ultrarelativistic limit of \req{eq:indchgF2} since $e\phi(0) \gg m_e$
\begin{equation}\label{eq:ultrarel}
\begin{split}
 m^2_{s,\mathrm{FD}}(\phi) &\approx \frac{4\alpha T^2}{3 \pi} \left[\pi^2 + \frac{\phi^2(r)}{T^2}\right]\,.
 \end{split}
\end{equation}
Which is simply the weak screening mass plus a strong screening correction. By removing the weakening screening portion, we get a strong screening mass that is independent of temperature
\begin{equation}
 m_{s,\mathrm{FD}}(\phi) \approx \sqrt{\frac{4\alpha}{3 \pi}} \phi_\text{vac}(0)\,,
\end{equation}
which describes the energy scale of the strong polarization potential at the origin. To get an exact value of the polarization potential at the origin, one could solve the zero temperature limit of \req{eq:PossBoltz} using the ultrarelativistic screening mass
\begin{equation}\label{eq:PossBoltz2}
 -\nabla^2\phi(r) + \frac{4\alpha}{3 \pi} \frac{\phi^3(r)}{(\hbar c)^2} = e\rho_{\mathrm{ext}}(r)/\varepsilon_0\,,
\end{equation}
which is essentially the ultrarelativistic limit of Thomas-Fermi theory or, in the case of a point charge, the Lane-Emden equation for $n=3$. This additive behavior also implies that calculations of the potential near the origin incorporating nonlinear screening at zero temperature can be added to estimates that calculate the long-distance thermal effects, such as Debye screening, without loss of accuracy. This is likely only the case in plasmas, which are globally weakly coupled, where weak screening is valid at large distances. 

The self-consistent strong potential found by solving \req{eq:PossBoltz} is very sensitive to boundary conditions and is only soluble when the form of the potential at large distances is known exactly. We expect that including additional gold atoms will change the self-consistent strong polarization potential contribution even at large ion separation distances.

\section{Enhancement factor for nuclear reaction rates}\label{sec:Enhancement}
We will assume that particles participating in reactions will be mostly at the Gamow energy \req{eq:gamow}. At this energy, the reacting light nuclei will penetrate up to
\begin{equation}\label{eq:turn}
 r_G(E_G) = \frac{Z\alpha \hbar c}{E_G}\,,
 \end{equation}
which is well into the screening cloud of the gold ion. Furthermore, if the particles are traveling on approximate lines of equipotential energy, they will experience a constant polarization potential over the tunneling integral in the screened cross-section
 \begin{equation}
 \sigma_\text{sc}(E) =\!\frac{S(E)}{E} \exp{\left( \! - \frac{2\sqrt{2 \mu_r}}{\hbar c}\!\!\int_{R}^{r_c} \!\!\! dr \sqrt{U_\text{sc}(r_G)\!- \! E}\right)}\,.
 \end{equation}
The screening enhancement factor, which is the ratio of the reaction rate $R$ in plasma to vacuum, for thermonuclear reactions with a constant polarization potential is the Salpeter enhancement factor\cite{Salpeter:1954nc}
 \begin{equation}\label{eq:weakenhance1}
 \mathcal{F}_{\text{sc}}\equiv \frac{R_\text{sc}}{R_\text{vac}} = \exp{\left( \frac{U_{vac}(r_G)-U_\text{sc}(r_G)}
 {T}\right)} 
 = \exp{\left( \frac{H(r_G)}
 {T}\right)} \, .
 \end{equation}
The potential $U_\text{sc}$ energy between the two nuclei is related to the potential $\phi$ calculated in Sect. \ref{sec:Solve} by the self-energy for n particles
\begin{equation}
 U_\text{sc}(r) = \frac{1}{2}\int d^3r' \rho(r') \phi(r-r') - U(r\rightarrow \infty) = \frac{1}{2}\int d^3r'\left[\sum^n_{i \neq j} \rho_i(r')\phi_j(r-r')\right] \,.
\end{equation}
We consider the self energies of the reacting nuclei and the heavy screening nuclei Where we have subtracted the in-plasma self-energies of the three nuclei separated at $r\rightarrow\infty$
\begin{equation}
 U(r\rightarrow \infty) = \frac{1}{2}\int d^3r'\left[\sum^n_{i=j}\rho_i(r')\phi_i(r-r')\right] \,.
\end{equation}
The total charge density is the sum of the induced plus the external charge density as in \req{eq:Poss}. We approximate $\rho_\text{ext}$ as a point charge at distances larger than $R$, relevant for the tunneling integral and neglect self energies at 2nd order in the polarization charge density $\rho_\text{ind}$. This will give us an underestimated change in the potential energy due to screening
\begin{equation}\label{eq:screen}
 U_\text{sc}(r) = \frac{1}{2}\left(Z_1 \phi_2 +Z_2 \phi_1 \right)+ \frac{Z_3}{2}\left(\phi_2 + \phi_1 \right)+ \frac{Z_1+Z_2}{2}\phi_3 \,.
\end{equation}
We choose $Z_1$ and $Z_2$ as the charges of the reacting particles and $Z_3$ as the heavy nucleus. The first term is the usual Salpeter enhancement factor but with a screening mass related to the electron density around the gold nucleus, the second term is the attraction of the gold atom to the polarization clouds of the reacting nuclei, and the final term is the attraction of the reacting nuclei to the polarization cloud of gold. 
\begin{equation}\label{eq:ratio}
 \mathcal{F}_{\text{sc}} = \exp{\left[\frac{1}{T}\left(Z_1 Z_2\alpha m_{s,\mathrm{FD}}(r_G) + Z_3 \frac{Z_1+Z_2}{2} \alpha m_{s,\mathrm{FD}}(r_G)+ \frac{Z_1+Z_2}{2}\phi_\text{ind}(r_G)\right)\right]}\,.
\end{equation}
The screening mass $m_{s,\mathrm{FD}}(r_G)$ is evaluated using the gold potential at $r_G$, representing the huge electron density around the gold nucleus. In our calculation, we will assume the turning point is the same for both reacting nuclei, which is only true for $Z_1 \approx Z_2 $. We will also assume the change in the tunneling probability will be small so that the Gamow energy does not change.

\begin{figure}%[h!]
\centering
\includegraphics[width=0.7\columnwidth]{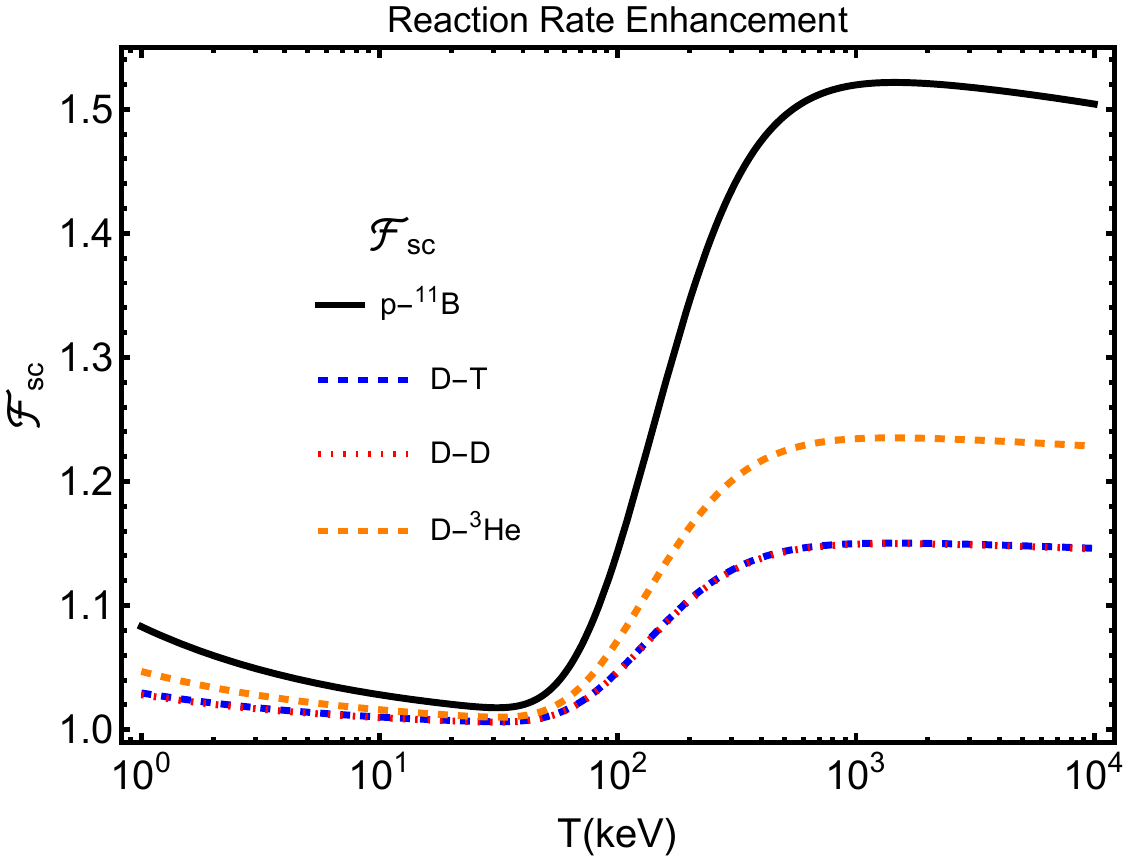}
 \caption{Reaction rate enhancement $\mathcal{F}_\text{sc}$ [see \req{eq:ratio}] for various fusion reactions as a function of temperature in keV: p-$^{11}$B as a black solid line, D-T as a dashed blue line, D-D as a dotted red line, and D-$^3$He as a dashed orange line.}
 \label{fig:enhance}
\end{figure}

Using these approximations, we arrive at the reaction rate enhancement shown in~\reff{fig:enhance}. The enhancement for each fusion reaction considered p-$^{11}$B, D-T, D-D, and D-$^3$He have similar temperature dependence. The slow rise at low temperatures is due to the strong screening contribution that is independent of temperature. The sharp rise above 100 keV is due to the thermal contribution to polarization becoming on the order of the strong screening contribution as seen in~\reff{fig:pot_Comp}. Also, the larger thermal kinetic energy at higher temperatures allows particles to penetrate further into the polarization cloud of gold. The enhancement flattens out at temperatures greater than the electron mass $m_e$ since there the electron polarization cloud is well described by the Debye screening mass~\reff{eq:debye}, which scales linearly with temperature in the ultrarelativistic limit, giving a constant enhancement factor.

Looking at each individual reaction, we can see that the largest effect on the enhancement is due to the size of $Z_1$ and $Z_2$. This is because a larger $Z$ more strongly attracts the local electron density to screen the reacting nuclei. This leads to p-$^{11}$B having the largest enhancement factor since boron has $Z_2=5$. The enhancement for each nuclear reaction is considerable and should be measurable in an experiment. Weak screening predicts much smaller changes to nuclear reaction rates~\cite{Hwang:2021kno,Grayson:2023flr}.

%%%%%%%%%%%%%%%%%%%%%%%%%%%%%%%%%%%%%
\section{Conclusion and discussion}\label{sec:Conc_Disc}

This work introduces the concept that fusion reactions may proceed more easily in the presence of ``heavy nuclear catalysts'' using a simple model of a heavy nucleus in an equilibrium plasma. Heavy catalysts are nuclei with a high atomic number ($Z$) that provide a `condensation' domain with a dense screening cloud of electrons. This environment allows higher energy light nuclei to undergo thermonuclear reactions with a significantly reduced Coulomb barrier by the high electron density.

We implemented self-consistent strong electron plasma screening near such a high atomic number nucleus to determine the short-range polarization potential relevant to thermonuclear reactions in an electron-ion plasma. We have in mind applications that arise, for example, when fusible material is found near such catalysts in the initial phases of laser interaction with gold nanorod-doped targets \cite{PRXEnergy.1.023001,universe9050233,csernai2023crater,csernai2024femtoscopy,szokol2024pulsedlaserintensitydependence,papp2024pics}.

We address the behavior of electron plasma at keV energies, which can be created through laser-matter interactions with intensities of $10^{17}-10^{20}$ W/cm$^2$ over timescales of roughly 20 femtoseconds. During this brief period, electrons in the plasma can achieve approximate thermal equilibrium. In contrast, the heavier ions remain immobile and do not contribute to the screening process other than by creating a location where many electrons coalesce. Using this picture, we use the equilibrium approach to solve the Poisson-Fermi-Dirac equation developed in \cite{Grayson:2024uwg}. A more complete plasma dynamics in self-consistent strong screening will be explored in future studies. The objective of this article is not to provide a definitive estimate of the enhancement of fusion reactions plasma but to improve on past works and to indicate that the effect of heavy nuclear catalysts could be measurable in experiments.

We comment on works by \cite{Wu:2016axg,Elsing:2022xyz} study similar strong screening enhancement in plasmas where an ion beam interacts with a cold electron gas, providing an electron density higher than the equilibrium density used here. We note that \cite{Wu:2016axg,Elsing:2022xyz} calculate much larger enhancements of thermonuclear reaction rates since they use a Boltzmann distribution for the electron density.

In \reff{fig:pot_Comp}, we determined the polarization potential $e\phi_{\text{ind}}$ for gold nuclei in an equilibrium electron-ion plasma at various temperatures. We find that fusion reactions of light nuclei occurring in the large electron cloud around a gold nucleus ($Z=79$) experience a keV scale lowering of the Coulomb barrier. 

The enhancement for p-$^{11}$B fusion is largest at temperatures above $100$ keV. At these temperatures, both the semi-bound and plasma screening electrons contribute. This effect is large enough, a factor $\sim 1.5$ enhancement, to be measured in future experiments of fusion reaction rates in plasma \cite{casey2023towards} and could be relevant for fusion in laser-generated fusion plasmas. This enhancement is most likely underestimated since we neglected contributions to the self-energy above linear order in the polarization charge~\req{eq:screen}, which may not be small. Moreover, enhancement is also predicted at low temperatures due to the bound gold electrons. The enhancement calculated here assumes reacting nuclei can only penetrate the electron cloud due to their thermal energy. If the reacting ions were non-thermal, for example, when generated by a (laser-accelerated) particle beam, they would penetrate much further into the electron cloud, yielding an enhancement rate on the order of
\begin{equation}
 \mathcal{F}_{\text{sc}} =\exp{\left(\frac{\langle Z\rangle e\phi_\text{ind}(R)}{T}\right)} \,,
\end{equation}
where $\langle Z\rangle$ is the average atomic number of the reactants and $e\phi_\text{ind}(R) \approx 14\,$keV. This would lead to even larger enhancement and could be relevant for measuring nuclear cross-section in metallic environments \cite{PhysRevC.78.015803}.

Historically, experimental determinations of astrophysical S-factors have found anomalous screening at low temperatures~\cite{SCHRODER1989466,Shoppa1993,ALIOTTA2001790,KCzerski2001}. This anomalous screening was measured recently~\cite{Zhang2020ApJ} again. The experimental results were never fully consistent regarding the screening magnitude, suggesting that some environmental properties were not fully controlled. In view of our results, we speculate the presence of target contamination by a catalyst, which at low temperatures achieved by intense beams could help explain anomalous screening at low collision energy. 

The enhancement from heavy nuclear catalysts increases with the average atomic number $\langle Z\rangle>2$ of the fusing nuclei, indicating catalysis is more beneficial for medium-$Z$ reactions, $\langle Z\rangle>2$, such as p-$^{11}$B. This is because such reacting nuclei more strongly attract local electron density, allowing stronger screening. The enhancement also increases with increasing catalyst atomic number $Z$ due to the increase in density of the electron screening cloud. %The effect of heavy nuclear catalysts can be further enhanced by increasing the electron density externally, such as with cold plasma jets.

Another interesting finding of this work is that the contribution to polarization potential at the origin from self-consistent strong screening is independent of temperature. This separation offers major simplifications to the determination screening enhancement of thermonuclear reactions. The self-consistent strong screening contribution to the potential can be estimated at zero temperature $e\phi/T \gg1 $ using the Thomas-Fermi model~\req{eq:PossBoltz2} or some equivalent model of the electron density around an atom. Then, thermal effects at large distances can be considered separately and summed up in the final result. In future studies, we will determine if this separation of thermal and strong screening also applies to a dynamic plasma.
 
In future theoretical work, we plan to improve our understanding of nuclear catalysts by addressing the limitations of our current model. A key simplifying assumption made to make our model soluble was equilibrium for electrons in the plasma. This is generally never true in laser-generated plasmas, which display mostly nonequilibrium behavior. A semi-analytical description of nonequilibrium effects through linear response may be possible in a combined dynamic-strong plasma theory. Additionally, while gold was treated as fully ionized in this study due to the high-temperature plasma context, in future calculations, we would like to describe the low-temperature quantum environment of the solid fusion target where conducting electrons contribute to the screening effect in a coherent response to the external field. 

We foresee an opportunity for experimental work in the near future dedicated to studying nuclear reactions in the presence of catalysts. This will help understand some ongoing screening enhancement riddles relevant to BBN and element genesis in stellar explosions. We can imagine a rather simple experimental setup for studying heavy catalysis of fusion reactions: An inverse kinematic experiment in which a heavy metal proton-hydride is the target for the Boron beam. Also, a slightly more sophisticated setup with a colliding beam p-$ {11}$B interaction point that lies within a thin gold foil. Such an experimental program is long overdue, seen the long-term riddle of screening effects in low-energy nuclear reactions.\\

\noindent \textbf{Acknowledgement:} \\
We thank Tamas Bir\'o and the Wigner Hun-Ren Research Center for their kind hospitality in Budapest during the PP2024 conference, supported by NKFIH (Hungarian National Office for Research, Development and Innovation) under awards 2022-2.1.1-NL-2022-00002 and 2020-2.1.1-ED-2024-00314. This meeting and the related research report motivate the presentation of these recently obtained results. 

CG gratefully acknowledges financial support for this research by the Fulbright U.S. Scholar Program, sponsored by the U.S. Department of State, the Hungarian Fulbright Commission, and the Hungarian Academy of Sciences under award FBM2024-2. Its contents are solely the author's responsibility and do not necessarily represent the official views of the Fulbright Program, the Government of the United States, or the Hungarian Fulbright Commission. JR did not receive sponsor funding for this work. \\

\noindent \textbf{Data Availability Statement:} \\
No datasets were generated or analyzed during the current study.
%%%%%%%%%%%%%%%%%%%%%%%%%%%%%%%%%%%%%%%%%%%%%%%%%%%%%%%%%%%%%%%%%.%
%===================================================================
%===================================APPENDICES======================
% \appendix
%===================================================================

\bibliographystyle{sn-aps} 
\bibliography{main}

\end{document}